\documentclass[sigconf]{acmart}

\usepackage{amsmath}
\usepackage{verbatim}
\usepackage{graphicx}
\usepackage{multirow}
\usepackage{array}
\usepackage{subfigure}
\usepackage{multicol}
\usepackage{color}
\usepackage{epsfig}
\usepackage{natbib}
\usepackage{xspace}
\usepackage{booktabs}
\usepackage{url}
\usepackage{amsfonts}
\usepackage{tabularx}

\copyrightyear{2021}
\acmYear{2021}
\setcopyright{acmcopyright}\acmConference[CIKM '21]{Proceedings of the 30th ACM
International Conference on Information and Knowledge Management}{November
1--5, 2021}{Virtual Event, QLD, Australia}
\acmBooktitle{Proceedings of the 30th ACM International Conference on Information
and Knowledge Management (CIKM '21), November 1--5, 2021, Virtual Event, QLD,
Australia}
\acmPrice{15.00}
\acmDOI{10.1145/3459637.3481926}
\acmISBN{978-1-4503-8446-9/21/11}
\newcommand{\hide}[1]{} 
\newcommand{\vpara}[1]{\vspace{0.1in}\noindent\textbf{#1 }}

\newcommand{\secref}[1]{Section~\ref{#1}} 
\newcommand{\figref}[1]{Figure~\ref{#1}} 
\newcommand{\tableref}[1]{Table~\ref{#1}}

\newcommand{\siamese}{Siamese\xspace}
\newcommand{\tablestrech}{0.8}
\newcommand{\methodshort}{HetMatch\xspace}
\newcommand{\bidword}{keyword\xspace}
\newcommand{\Bidword}{Keyword\xspace}
\newcommand{\bidwords}{keywords\xspace}
\newcommand{\bidpair}{$<$ad, keyword$>$\xspace}
\newcommand{\zhitongche}{Alibaba's e-commerce sponsored search platform\xspace}

\newcolumntype{L}[1]{>{\raggedright\arraybackslash}p{#1}}
\begin{document}
\fancyhead{}
\title{Heterogeneous Graph Neural Networks for Large-Scale \\Bid \Bidword Matching}
\hide{
\author{Zongtao Liu}
\affiliation{
  \institution{Alibaba Group}
   \country{China}
}
\email{zongtao.lzt@alibaba-inc.com}

\author{Bin Ma}
\affiliation{
  \institution{Alibaba Group}
   \country{China}
}
\email{mabin.mb@alibaba-inc.com}

\author{Quan Liu}
\affiliation{
  \institution{Alibaba Group}
   \country{China}
}
\email{lq204691@alibaba-inc.com}

\author{Jian Xu}
\affiliation{
  \institution{Alibaba Group}
   \country{China}
}
\email{xiyu.xj@alibaba-inc.com}

\author{Bo Zheng}
\affiliation{
  \institution{Alibaba Group}
   \country{China}
}
\email{bozheng@alibaba-inc.com}
}
\author{
	Zongtao Liu, Bin Ma, Quan Liu$^{*}$, Jian Xu, Bo Zheng} \thanks{$^*$Corresponding author: Quan Liu, lq204691@alibaba-inc.com}
\affiliation{Alibaba Group \country{China}}
\email{{zongtao.lzt, mabin.mb, lq204691, xiyu.xj, bozheng}@alibaba-inc.com}

\renewcommand{\shortauthors}{}


\begin{abstract}
Digital advertising is a critical part of many e-commerce platforms such as Taobao and Amazon. While in recent years a lot of attention has been drawn to \textit{the consumer side} including canonical problems like ctr/cvr prediction, \textit{the advertiser side}, which directly serves advertisers by providing them with marketing tools, is now playing a more and more important role. When speaking of sponsored search, bid keyword recommendation is the fundamental service. This paper addresses the problem of keyword matching, the primary step of keyword recommendation. Existing methods for keyword matching merely consider modeling relevance based on a single type of relation among ads and keywords, such as query clicks or text similarity, which neglects rich heterogeneous interactions hidden behind them. To fill this gap, the keyword matching problem faces several challenges including: 1) how to learn enriched and robust embeddings from complex interactions among various types of objects; 2) how to conduct high-quality matching for new ads that usually lack sufficient data.

To address these challenges, we develop a heterogeneous-graph-neural-network-based model for keyword matching named HetMatch, which has been deployed both online and offline at the core sponsored search platform of Alibaba Group. To extract enriched and robust embeddings among rich relations, we design a hierarchical structure to fuse and enhance the relevant neighborhood patterns both on the micro and the macro level. Moreover, by proposing a multi-view framework, the model is able to involve more positive samples for cold-start ads. Experimental results on a large-scale industrial dataset as well as online AB tests exhibit the effectiveness of HetMatch. 
\end{abstract}

\maketitle

\section{Introduction}
\label{sec:intro}
\begin{figure}[!h]
	\centering

	\includegraphics[width=0.44\textwidth]{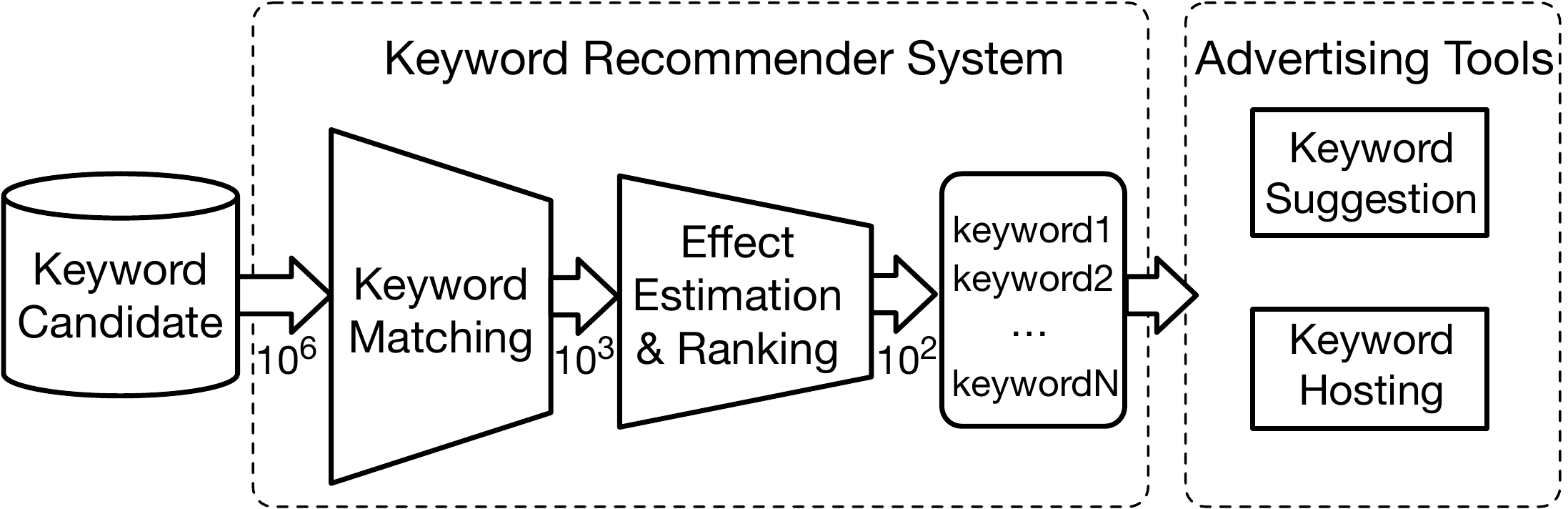}
	\caption{A real-world integrated \bidword recommender system.}
	\label{fig:recommender}
	\vspace{-0.1in}
\end{figure} 

Sponsored search is one of the main fashions of online advertising where advertisers acquire their desired ad impressions or clicks via bidding proper \bidwords. At the sponsored search platform of Alibaba, millions of advertisers in total manually add tens of millions of \bidwords every day, which reflects advertisers' strong bidding willingness to make their products obtain more potential consumers. Compared with such strong willingness, many advertisers lack expert knowledge to choose proper \bidwords thus fail to get expected ad impressions or clicks for their products.  As illustrated in the previous empirical study \cite{zhang2014bid}, many advertisers tend to bid on a small number of popular keywords in the advertising platform, which makes those advertisers with low bid prices harder to obtain impressions. This challenge is also common at the sponsored search platforms in Alibaba, where only less than 10\% of handcrafted selected \bidword can obtain ad impressions on the next day. To improve marketing effectiveness and lighten the burden of advertisers, many sponsored search platforms are now offering different products that perform \bidword recommendation, either in a white-box or a black-box fashion. White-box products offer many \bidword candidates where advertisers can manually choose from, while black-box tools provide automatic \bidword hosting services. 
\hide{
Recommender systems (RS) in the field of online advertising or e-commerce is usually seen as a personalized advertisement or item recommendation. It attempts to predict users' preference for given products based on profiles, behaviors, context, etc. On the other hand, which connects the merchants or advertisers, recommender systems can be equally useful and important, though often less attention is drawn to. Without effective merchant tools that transform advertisers' bidding willingness into concrete bidding \bidwords, prices, and offers, nothing could be retrieved, ranked, or presented when serving online consumers. For sponsored search advertising, advertisers choose concrete bid \bidwords among sea of words, which later get matched to queries that consumers search when surfing. Traditionally, advertisers often manually select various bid \bidwords for each ad based on their domain knowledge and marketing experience, which is time-consuming and usually not effective. To lighten such burden of advertisers, advertising platforms are now offering different products that perform \bidword recommendation, either in a white-box or a black-box fashion. White-box products offer many \bidword candidates where advertisers manually choose from, while black-box tools provide automatic \bidword hosting services. 
}

Like many other industrial recommender systems (RSs), an advertising \bidword recommender system can be achieved via two phases: 1) \bidword matching/retrieval which collects a subset of tokens among sea of words for a given ad; 2) efficient ranking which puts an order on individual subset based on relevance and estimated effect (C.f. \figref{fig:recommender}).
In this work, we address the problem of \bidword matching, which is a fundamental step of \bidword recommendation and also plays a determining role in the quality of the ranking stage.

\begin{figure}[t]
	\centering
	\subfigure[Network schema] {
		\includegraphics[width=0.20\textwidth]{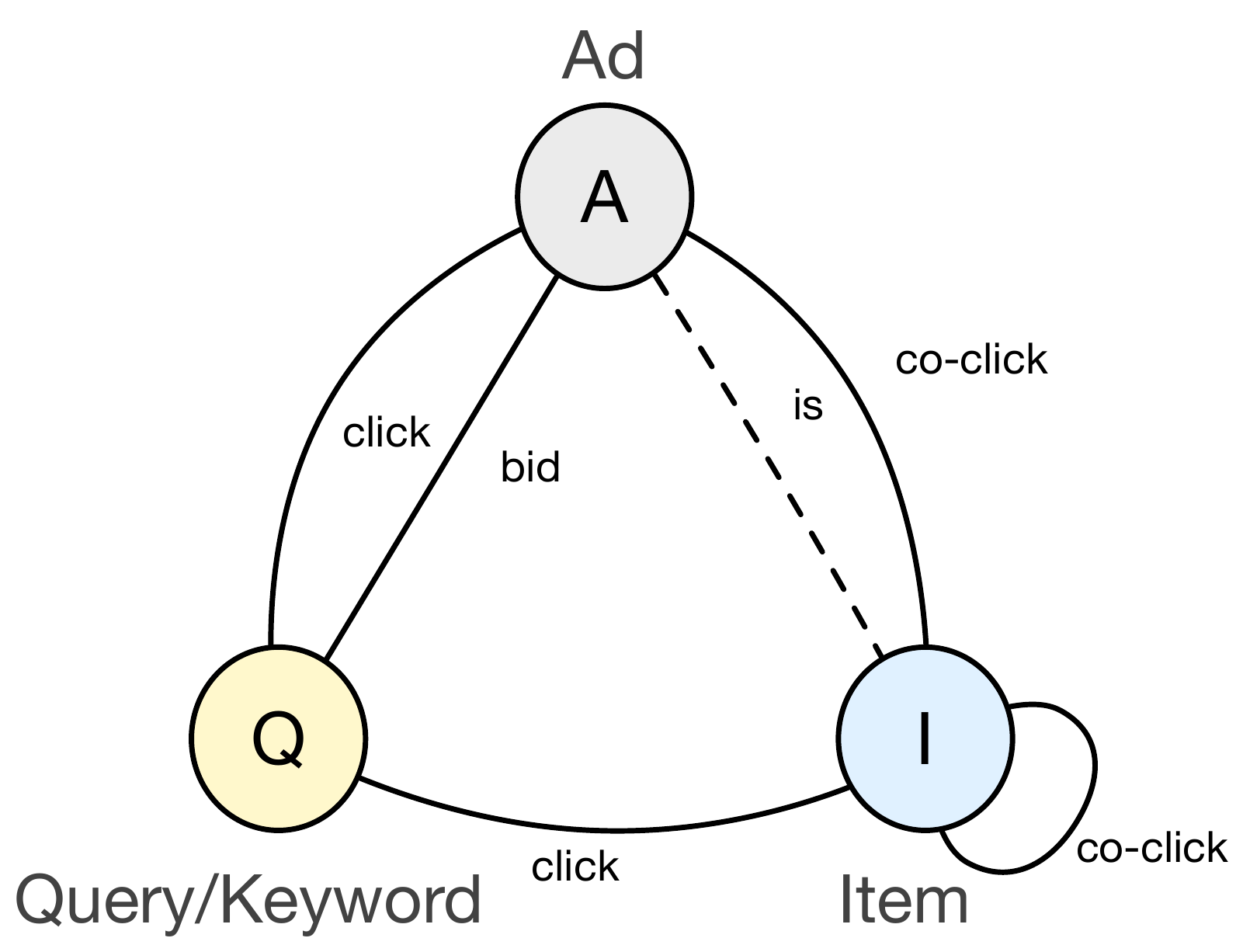}
	}
	\subfigure[Heterogeneous information network] {
		\includegraphics[width=0.24\textwidth]{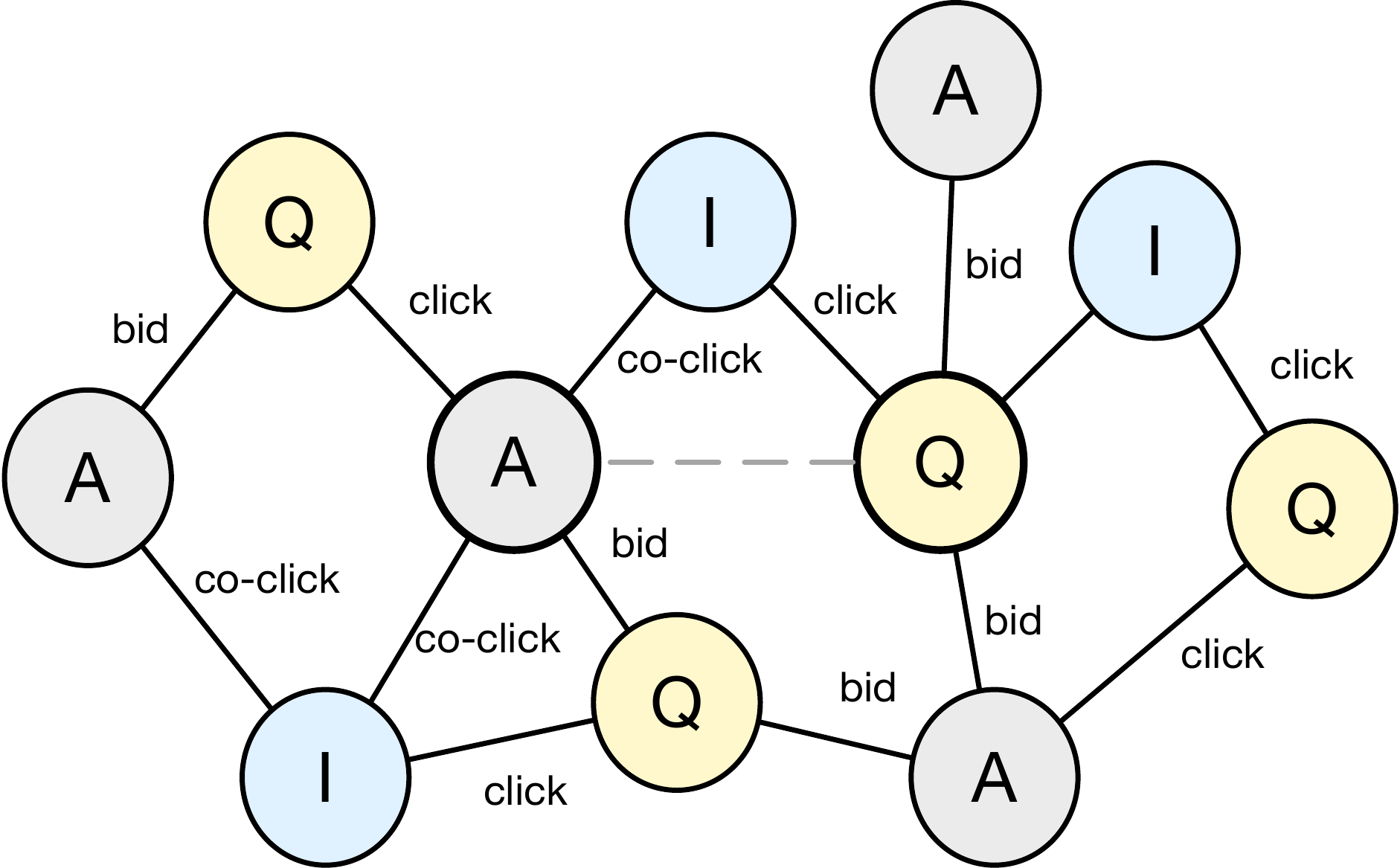}
	}
	\hfill
	\caption{HIN schema for e-commerce bid \bidword matching problem. An \textit{ad} refers to an ad-group created in the advertising platform,  a \bidword can either denote a bidding keyword or an ordinary  query searched by users, and an \textit{item} denotes an ordinary goods in the e-commerce platform. Besides, each ad is corresponding to a unique item.}
	\label{fig:schema}
\end{figure}

Regarding the \bidword matching or retrieval problem, various approaches have been explored over years, including text similarity, collaborative filtering, and topic-based ones \cite{bartz2006logistic,joshi2006keyword}. However, there are several limitations of existing methods: 
1) these approaches neglect rich heterogeneous information hidden behind the simple \bidpair pair (\figref{fig:schema} illustrates a toy example of HIN schema in which ads, \bidwords, and items are denoted by different types of nodes), which helps understanding the relevance between them ; 
2) This also causes the cold-start problem to be worse for new ad units, especially those of new advertisers. 
\hide{
\begin{figure}[t]
	\centering
	 \hspace{-0.3in}
	\subfigure[Noise filtering on the micro level] {
		\includegraphics[width=0.13\textwidth]{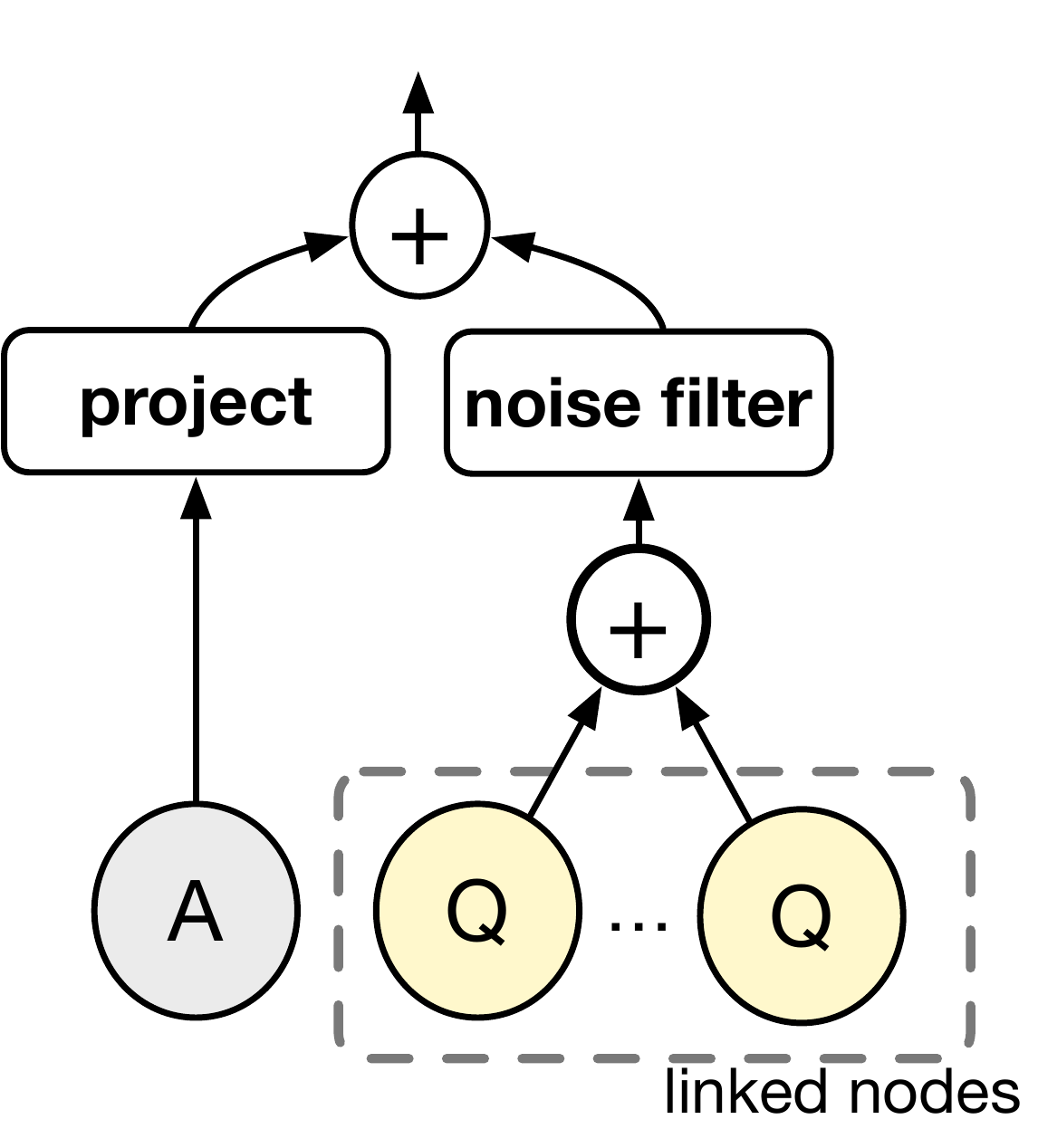}
	}
	\hspace{0.1in}
	\subfigure[\siamese neighbor matching on the macro level] {
		\includegraphics[width=0.13\textwidth]{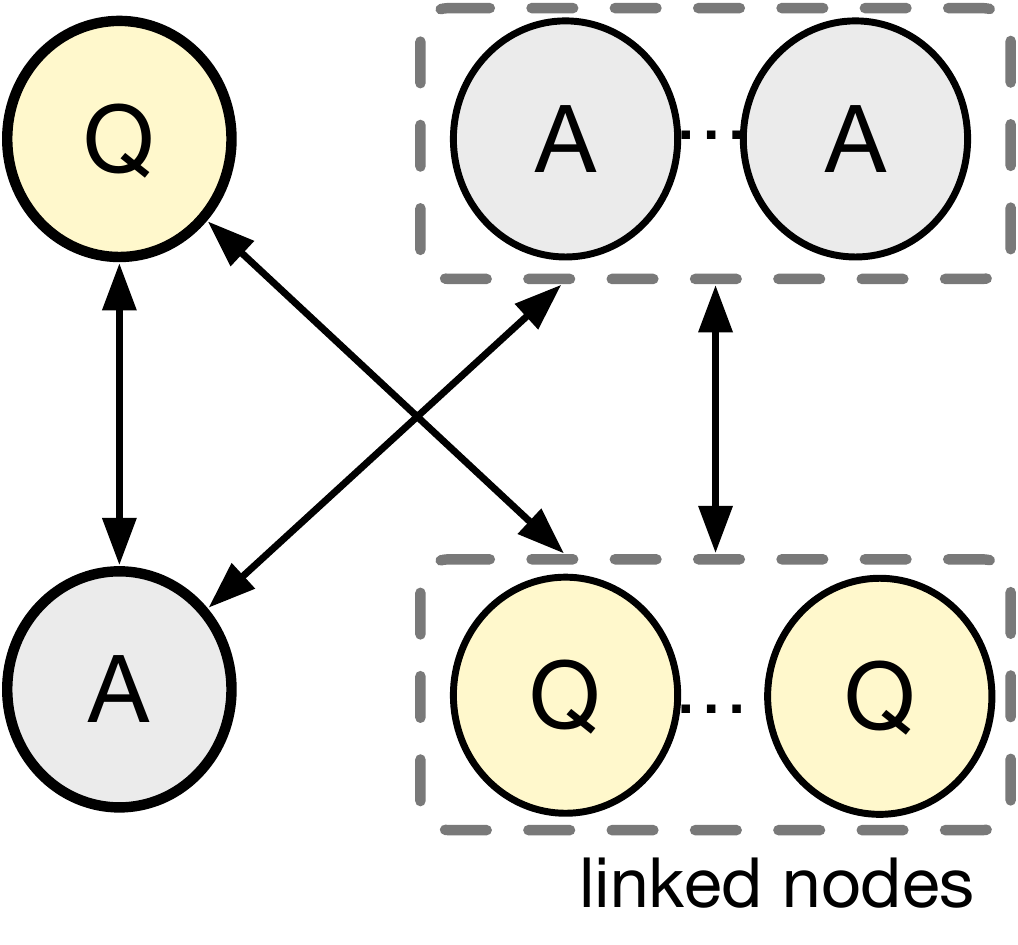}
	}
	\hspace{0.1in}
	\subfigure[Multi-view objectives for alleviating cold-start problem] {
		\includegraphics[width=0.13\textwidth]{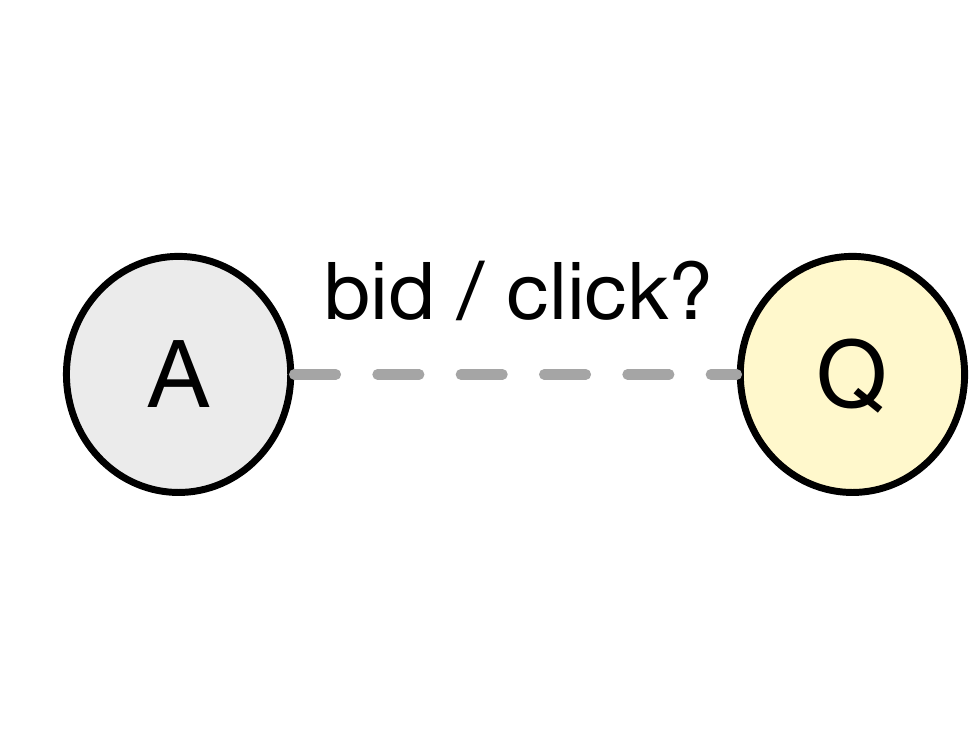}
	}
	\hfill
	\caption{Overview of our proposed framework.}
	\label{fig:framework_overview}
\end{figure}
}
\hide{
\begin{figure}[t]
	\centering
	\includegraphics[width=0.48\textwidth]{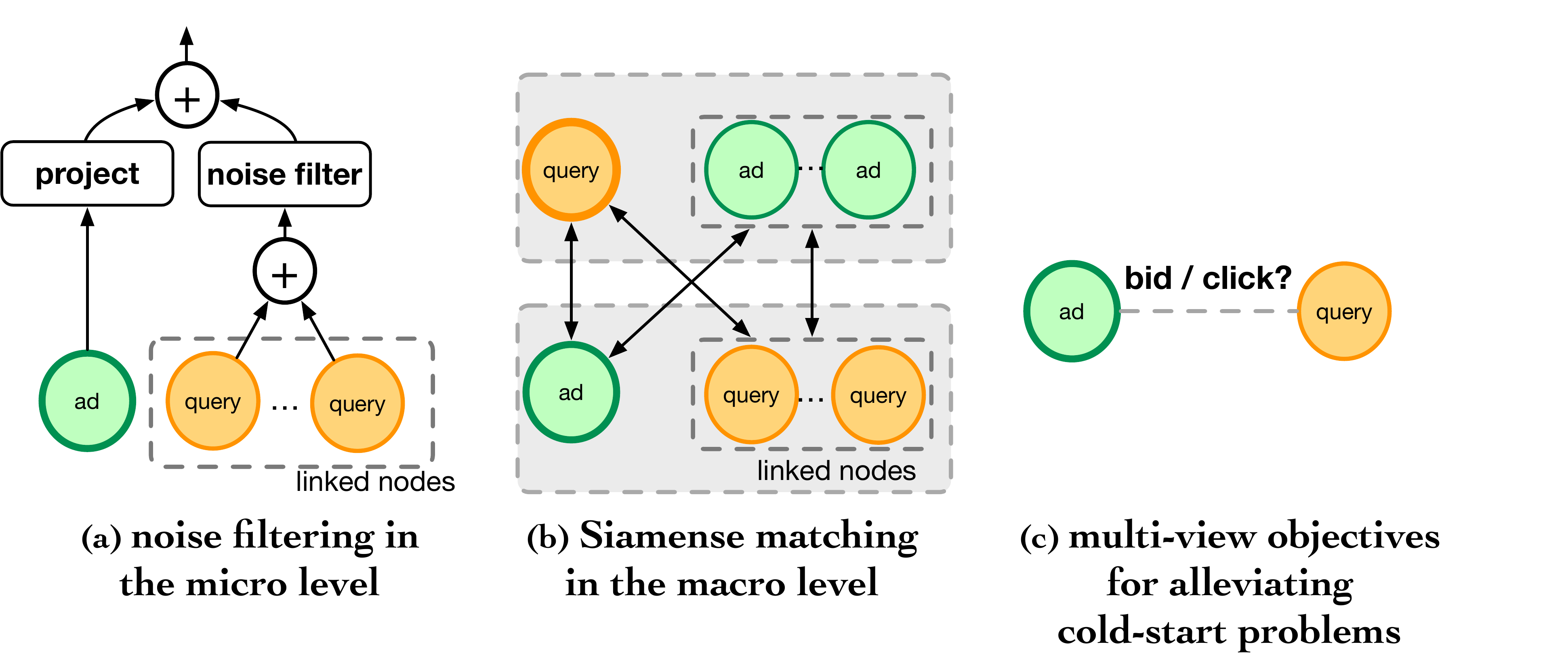}
	\caption{overview of our proposed framework}
	\label{fig:framework}
\end{figure} 
}

To utilize the rich information hidden behind ads and \bidwords, we resort to heterogeneous graph neural networks (HGNNs), which have various successful applications in modeling complex interactions between different types of objects \cite{dong2017metapath2vec,fu2017hin2vec}, and then integrate HGNNs in an embedding-based framework which is proven effective in multiple information retrieval tasks \cite{shi2016survey}.
Embedding-based methods have been extensively studied in recommendation scenarios to improve the quality of matching. These methods aim to represent each node with different types of features in heterogeneous information network as
a low-dimensional embedding vector, expecting that similar source objects (ads) and target objects (keywords) have similar embeddings. Among them, GNN-based approaches benefit from their strong ability to fuse relevant information from neighbors at different distances in the network, thus achieving state-of-the-art performance in matching task \cite{DBLP:journals/corr/abs-2011-02260}. More recent studies have extended GNN to heterogeneous information networks to capture the complex interactions among various types of nodes\cite{shi2016survey}.
 Most of these studies achieve this goals by aggregating neighbors' information by sampling subgraphs via different metapaths \cite{DBLP:conf/www/ZhongLAHFTH20,DBLP:conf/kdd/NiuLLXSDC20}, where a metapath denotes a sequence of meta relations (C.f. \secref{sec:setup}). 
 
 For heterogeneous recommendation such as keyword recommendation,  the  selected metapaths vary a lot between different sides of a two-tower structure \cite{DBLP:conf/kdd/NiuLLXSDC20, DBLP:conf/www/ZhongLAHFTH20}, where a tower refer to a sub-network to compute a node's final embedding, resulting that the sampled attributed subgraph in each tower is very different. More specifically, the types of nodes, edges and their corresponding features in the same position of each subgraph vary a lot. In this way, it might be hard to guarantee that the final embeddings of two towers  fall into adjacent feature space. Besides, it is also a challenging problem to filter irrelevant information when involving heterogeneous relations and attributes. This problem is especially severe in industrial web-scale scenes such as in \zhitongche. How to learn 
comprehensive and robust embeddings in modeling complex interaction among nodes is an intractable problem.

\underline{}
Another factor affecting the quality of \bidword matching is the cold start of new ads. In Taobao, a leading e-commerce platform in China, millions of new ads are continuously created by advertisers every day. However, there are no user behaviors for these new ads throughout the platform. How to process these ``cold start"  ads is another challenging problem. 

Motivated by the concerns mentioned above, this paper  proposes a \textbf{Het}erogeneous graph neural network for \bidword \textbf{Match}ing (\methodshort) for e-commerce sponsored search. To enhance the robustness of the extracted embeddings with various semenatics, on the macro level, we develop a \siamese network architecture by aggregating each tower's final embeddings and the averaged neighborhood embedding to conduct matching. In this way, the original heterogeneous matching problem between ads and \bidwords is transformed into a homogeneous matching between two \bidpair pairs. 
On the micro level, we apply autoencoder in graph convolution as a fundamental component of our model to reduce the influence of noisy signals and the computational cost. Moreover, to improve the capability to match cold-start ads, we leverage a multi-view framework to use multiple types of relations besides click data of ads as our objectives.

Accordingly, our contributions are as follows:
\begin{itemize}
	\item We propose a novel HGNN-based \bidword matching framework that leverages the complex relational data behind ads and keywords. To extract enriched and robust embeddings among different relations, on the micro level, we apply an autoencoder in graph convolution to mitigate the irrelevant patterns in neighborhood; while on the macro level, we develop a \siamese neighbor matching layer on the top of HGNNs to involve more relevant neighborhood information. 
	\item To introduce more supervised signals for alleviating the cold-start problem, we use a multi-view framework to learn the posterior probability of various objectives. 
	\item Experiment results exhibit our model's advantages over other methods both on offline and online evaluation.
	\item To the best of our knowledge, it is the first work to apply a HGNN-based method to \bidword recommendation, which consolidates the foundation of sponsored search.
\end{itemize}

\hide{
\vpara{Organization.} The remainder of this paper is organized as follows. We first formulate the problem and review some relevant research. We then give necessary definitions and introduce the proposed approach, including the design and motivation of each component. Subsequently, we present the experimental results and conclude the work.
}

\section{Related Work}
\label{sec:related}

\vpara{\Bidword Recommendation and Suggestion.}
Early methods of bid \bidword recommendation in sponsored search mainly focus on retrieving \bidwords from the perspective of relevance. These methods can be broadly categorized into the following types: collaborative methods \cite{bartz2006logistic},  proximity-based methods \cite{joshi2006keyword,abhishek2007keyword}, and topic-based methods \cite{chen2008advertising}. Some collaborative methods like \cite{bartz2006logistic} aim to mine phrases that co-occur with the seed queries, which is applied in Google's Adword Tools. Proximity-based approaches mainly focus on designing different distance metrics to evaluate the similarity between queries, which can be further divided into network-based methods \cite{joshi2006keyword} and kernel-based methods \cite{abhishek2007keyword}. Topic-based methods use a hierarchical or unsupervised approach to cluster queries to topic \cite{chen2008advertising}. These methods mainly address the relevance of retrieved corpus and ignore modeling the effect of selected \bidwords; thus may not perform well in practice.  

Another line of \bidword recommendation focuses on retrieving the \bidwords from click-logs and estimate the user-click or impression brought by selected \bidwords \cite{fuxman2008using,nath2013ad,wang2009search}.
For instance, \citet{fuxman2008using} propose algorithm within Markov Random Field model to estimate user clicks. 
However, these methods are computationally expensive and lack the generalization ability for new ads; thus are usually applied to ranking retrieved \bidwords rather than performing matching tasks.


\vpara{Representation Learning on Heterogeneous Networks.} Network representation learning (NRL) aims to automatically encode node information and network structure to fixed-sized latent representation, which can be used in downstream graph mining tasks. Over the years, quite a few NRL methods have been proposed. These methods can be divided into random-walk-based methods \cite{dong2017metapath2vec},  factorization-based methods \cite{qiu2018network} and 
GNN-based methods \cite{DBLP:conf/esws/SchlichtkrullKB18,DBLP:conf/kdd/ZhangSHSC19,DBLP:conf/www/WangJSWYCY19}. Earlier works mostly focused on studying the network representation learning method on homogeneous graphs. Recently, inspired by the past homogeneous NRL, studies have attempted to extend homogeneous NRL methods on heterogeneous networks. And the core problem is to fuse diverse node attributes and different types of relations into a latent vector; thus, the extracted embeddings can preserve both semantic and structural properties among nodes. Based on previous random-walk-based homogeneous NRL, \citet{dong2017metapath2vec} and \citet{fu2017hin2vec} introduce metapath-based random walk strategies, in which the walker is confined to transit based on given metapaths. 
On the other line of works, applying GNNs on heterogeneous networks has also been proven a promising direction \cite{DBLP:conf/esws/SchlichtkrullKB18,DBLP:conf/kdd/ZhangSHSC19,DBLP:conf/www/WangJSWYCY19}. 
For instance, \citet{DBLP:conf/www/WangJSWYCY19} maintains different weights for different metapath-defined edges when applying aggregation in the graph attention networks. However, current HGNN-based methods do not explicitly address the importance of filtering irrelevant patterns propagated through complex interactions, which is achieved by our extended version of graph convolution.

\vpara{Heterogeneous-Information-Network based Recommendation.}
More recently, leveraging heterogeneous information networks (HIN) becomes an emerging direction in recommender systems due to its capability of characterizing complex objects and rich relations \cite{shi2016survey}.
For instance,
\citet{feng2012incorporating} proposed to alleviate the cold start issue with heterogeneous information network contained in the social tagged system.
Metapath-based methods were introduced into hybrid recommender system in \cite{yu2013recommendation}.
Recently, \citet{hu2018leveraging} leveraged metapath-based context in top-N recommendation.
However, none of them addresses the problem of the heterogeneity of metapaths between two towers, which might lead to difficulties for learning a robust embedding for matching with very different computational graphs.

\section{Problem Setup}
\label{sec:setup}

Recently, several researches have  explored how to utilize the rich relations and complex objects of heterogeneous information network (HIN) in recommendation systems, which is an auspicious direction. \Bidword recommendation also deals with various types of objects and rich relations (C.f. \figref{fig:schema}), thus it would be promising to leverage the HIN-based recommendation paradigm. Here we first give the necessary concepts about HIN.
\begin{definition}{\textbf{ Heterogeneous Information Network} \cite{DBLP:journals/sigkdd/SunH12}.}
	A heterogeneous information network (HIN) can be denoted as  $\mathcal{G} =(\mathcal{V},\mathcal{E}, \phi, \psi)$, consisting of an object set $\mathcal{V}$ and a relation set $\mathcal{E}$.  Each node $v$ owns a type $\phi(v)$, and each edge $e$ has a type $\psi(e)$.
	$\mathcal{T}$ and $\mathcal{R}$ denote 
	the sets of predefined object types and relation types, where $|\mathcal{T}|+|\mathcal{R}|>2$. $\mathcal{V}$ can be written as $\mathcal{V}=\mathcal{V}_{t_1}\cup...\cup \mathcal{V}_{t_i} \cup...\cup\mathcal{V}_{t_{|T|}}$, where $t_i\in \mathcal{T}$ and $\mathcal{V}_{t_i}$ represents the node set with type $t_i$. Besides, each node $v_t \in \mathcal{V}_t$ is augmented with node attribute $\mathbf{x}^t\in \mathbb{R}^{d_t}$, where $d_t$ is the attribute dimension of the nodes typed $t$. Similarly, $\mathcal{E}$ can be written as $\mathcal{E}_{r_1}\cup...\cup\mathcal{E}_{r_i}\cup...\cup\mathcal{E}_{r_{|\mathcal{R}|}}$, where $r_i \in \mathcal{R}$ and $\mathcal{E}_{r_i}$ represents the edge set with type $r_i$. 
\end{definition}
More specifically, in the scenario of \bidword recommendation,  we denote the whole node set $\mathcal{V}=\mathcal{V}_A\cup\mathcal{V}_Q\cup\mathcal{V}_I$, where $\mathcal{V}_A$, $\mathcal{V}_I$ and $\mathcal{V}_Q$ represents the ad set, \bidword set, and item set, respectively. More specifically, an \textit{ad} refers to an ad-group created in the advertising platform,  a \bidword can either denote a bidding keyword or an ordinary  query searched by users, and an \textit{item} denotes an ordinary goods in the e-commerce platform. Among these nodes, there exists various types of relations which capture different semantic connections, such as the bid or click relations between ads and keywords.  
In order to capture the semantic and structural relation between different objects, the metapath is proposed by \cite{DBLP:journals/pvldb/SunHYYW11} as a relation sequence connecting two objects, and is widely applied in HIN model research.
\begin{definition}{\textbf{Metapath and Metapath-guided Neighbors} \cite{DBLP:journals/pvldb/SunHYYW11}.}
	For a relation $e=(s,t)$ linking from node $s$ to $t$, its meta relation is defined as $<\phi(s),\psi(e),\phi(t)>$ . A meta-path $p$ is then defined as a sequence of meta relations $t_1 \xrightarrow{r_1} t_2 \xrightarrow{r_2} \cdots \xrightarrow{r_l} t_{l+1}$, which describes a composite relation $R = r_1 \circ r_2 \cdots \circ \cdots  r_l$ between nodes $a_1$ and $a_{l+1}$, where $\circ$ denotes the composition operator on relations. Furthermore, the metapath-guided neighborhood is defined as the set of all visited nodes when a given node $v$ walks along the given metapath $p$. 
\end{definition}

Finally we present the problem definition of HIN-based \bidword matching.
\begin{definition}{\textbf{ HIN-based Ad-\Bidword Matching Problem.}}
	Given a HIN $\mathcal{G} =(\mathcal{V},\mathcal{E})$,  let $\mathcal{A}=\{a_1, a_2, ..., a_N\} \subset \mathcal{V_{A}}$  be a set of selected ads to match \bidwords, and $\mathcal{E}^c, \mathcal{E}^t\subset \mathcal{A}\times\mathcal{V}_\mathcal{Q}$ be the candidate relation set and target relation set respective, where `$\times$' represents the Cartesian product operator between two set. 
	More specifically, each ad $a_i$ corresponds to a target \bidword set $\mathcal{Q}_i^t \subset \mathcal{V}_\mathcal{Q}$ and a candidate \bidword set $\mathcal{Q}_i^c \subset \mathcal{V}_\mathcal{Q}$.
	
	The ad-\bidword matching problem can be formulated as a recall optimization task \cite{huang2020embedding}. Given a set of ads $\mathcal{A}=\{a_i\}_{i=1}^N$ with their corresponding candidate relation sets $\mathcal{E}^c=\bigcup_{i=1}^N(a_i\times\mathcal{Q}_i^c)$ and the target relation sets $\mathcal{E}^t=\bigcup_{i=1}^N(a_i\times\mathcal{Q}_i^t)$, the problem is to pick a retrieved  relation set $\mathcal{E}^o=\bigcup_{i=1}^N(a_i\times\mathcal{O}_i^{K})$, where $\mathcal{O}_i^{K}$ represents the retrieved \bidwords for ad $a_i$ sized no more than $K$, the object is to maximize the total recall ratio:
	\begin{eqnarray}
	Recall@K = \frac{\sum_i^N{|\mathcal{O}_i^K\cap \mathcal{Q}_i^t|}}{\sum_i^N|\mathcal{Q}_i^t|}
	\end{eqnarray}
	Usually, the candidate set of a matching problem is set as a group of target  nodes which is related to the source nodes under a certain criterion. In our task, we define the candidate set $Q_i^c$ of an ad $a_i$ as the keywords that share the same category with $a_i$, which is adjusted by a trained category classifier. 
	The target relation set $\mathcal{E}^t$ contains the node pair of \bidpair which would bring user clicks in the future.
	In other words, we aim to optimize our matching model in that the matched \bidwords can cover as many effective bidding \bidwords as possible. 
\end{definition}

\section{Method}
\begin{figure*}[t]
	\centering
	\includegraphics[width=0.8\textwidth]{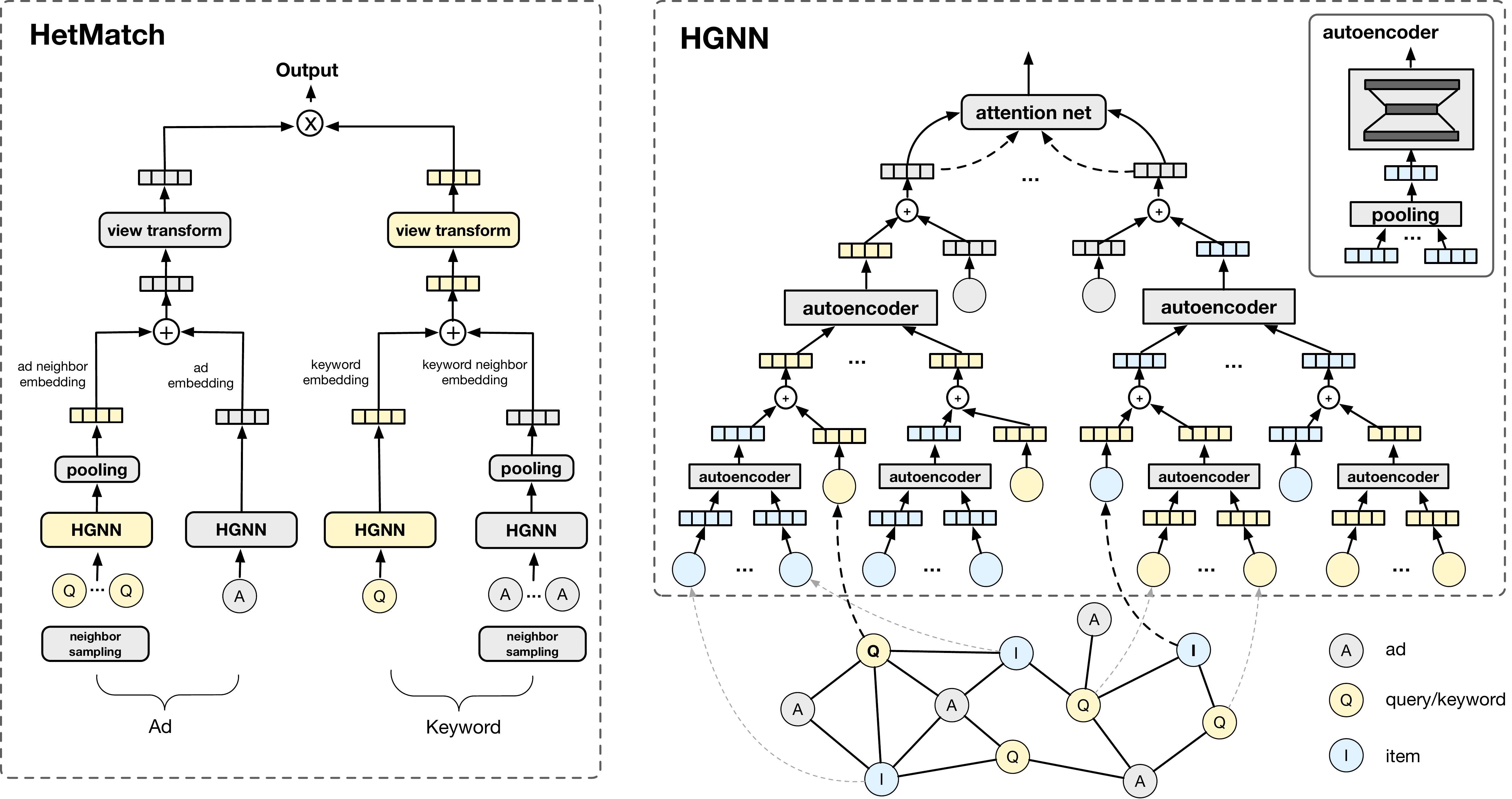}
	\vspace{-0.02in}
	\caption{Architecture of our proposed model. The left part is the two-tower structure of our proposed method in which we use a \siamese neighborhood matching layer to involve more homogeneous node patterns and utilize a view-transformation module to involve more supervised signals. The right part is the metapath-based heterogeneous GNN in which we use an autoencoder to reduce  the influence of noisy signals and the computational cost.}
	\label{fig:model}
	\vspace{-0.02in}
\end{figure*}

In recent years, attention in the field of RS is increasingly shifting towards HIN-based recommendation. We extend this paradigm of RS for bid \bidword matching problem and design a hierarchical network architecture to model the enriched interactions behind ads and keywords in a robust way. On the micro level, we leverage metapath-based graph convolution to aggregate neighborhood context from different perspectives and utilize an autoencoder module to filter irrelevant neighborhood patterns. On the macro level, we design a \siamese network structure to enhance the complementary patterns in the neighborhood. To further alleviate the cold start problem of newly created ads, we extend our framework as a multi-view matching problem among different relation targets between ads and \bidwords. In this way, more supervising information are introduced and the associated tasks can improve each other's performance by sharing information.
In this section, we describe the details of our proposed model, \textbf{Het}erogeneous Graph Neural Network for \bidword \textbf{Match}ing (\methodshort). \figref{fig:model} shows the architecture of our proposed model.
\subsection{Node-Level Fusion}
Before integrating data from metapath neighbors for each node, 
we first extract heterogeneous feature vector $\textbf{x}^t\in \mathbb{R}^{d_t}$ from $\mathcal{X}^t$ for each node $v_t\in \mathcal{V}_t$ and transform it as a fixed-length embedding. We list the details of features in \tableref{tab:feature}. These features can be categorized into two types, i.e., id features and numeric features. We transform each numeric feature as a discrete value, which represents the quantile of its feature distribution. After discretization, we can acquire the embedding of each feature via its corresponding look-up table. To lighten the computation burden, the same features from different node types will share one look-up table. For instance, the feature \textit{terms of title} appears both in ad's and item's feature list, and thus the embedding of this feature shall be acquired from the same look-up table. We then concatenate these embeddings and feed the concatenation into a type-specific neural network $f_t$ to get the node-level embedding $\mathbf{h}_{v_t}$.  

\begin{table}[!hpb]
	\centering
	\small
	\caption{List of features in this paper.}
	\renewcommand{\arraystretch}{\tablestrech}
	
	\begin{tabular}{ll}
		\toprule
		Node Type&  \multicolumn{1}{l}{Features} \\
		\midrule
		ad &  \multicolumn{1}{p{6.5cm}}{ad id, terms of title/description, category path, brand, properties and shop information}  \\
		\midrule
		item &  \multicolumn{1}{p{6.5cm}}{item id, terms of title/description, category path, brand, properties and shop information}\\
		\midrule
		keyword & \multicolumn{1}{p{6.5cm}}{keyword id, terms of keyword, predicted category, average bid price of the keyword, the number of bid on the keyword, the total in-shop count led by the keyword}\\
		\bottomrule
	\end{tabular}
	\label{tab:feature}
\end{table}

\subsection{Subgraph-Level Fusion}
\label{sec:subgraph}
\vpara{Metapath.}After fusing the node-level information, the next step is to involve the rich context information around each node. Under the paradigm of metapath-based heterogeneous GNNs, multiple metapaths are sampled which play different roles in capturing the structural and semantic context. We thus introduce the metapaths used in our model, which can be categorized into the following two groups:

(1) \textit{Bid-based group}: The subgraphs based on bid relations around a given ad or \bidword can directly characterize the bidding environment around it,  which involves the competitive ads that bid on the shared \bidwords and a surge of \bidwords that the competitive ads are interested in. We use the following four metapaths constructed by user clicks and advertiser bids to capture such environment:

\begin{itemize}
	\item $q\xrightarrow{click} a \xrightarrow{click}$q, \quad $q\xrightarrow{click} a \xrightarrow{bid}$q
	\item $a\xrightarrow{click} q \xrightarrow{click}$a, \quad $a\xrightarrow{bid} q \xrightarrow{click}$a
\end{itemize}

\noindent where the user-click relations can stand for the bids that can directly bring clicks, and the ordinary bid relations can make appropriate supplements to the cold-start ads.

(2)\textit{Item-based group}:  Sometimes a user might pay attention to an ad and an ordinary item in the same page view. These items can build bridges between ads and related \bidwords that are not directed connected and provide extra useful semantics that helps capture richer context information. Besides,  those co-clicked items can also offer similar textual and behavioral patterns that can characterize how the connected nodes are like. 
\begin{itemize}
	\item $q\xrightarrow{click} i \xrightarrow{co-click}$a
	\item $a\xrightarrow{co-click} i \xrightarrow{click}$q
\end{itemize}

\vpara{Graph Convolution Layer with AutoEncoder.} After introducing the metapaths used in our model, we then present how we aggregate the node-level embeddings robustly and efficiently given a certain metapath. The core idea of graph neural networks is to iteratively aggregate feature information in the neighborhood by a local filter. However, the neighborhood information aggregated by different heterogeneous relations might contain irrelevant information. Without loss of generality, we apply the autoencoder, which is successfully used in previous tasks for learning the compressed representation for denoising \cite{DBLP:conf/icml/VincentLBM08}, to filter irrelevant neighborhood patterns and meanwhile improve the efficiency of aggregating. The process of graph convolution can be formulated as below:
\begin{eqnarray}
h_{\mathcal{N}_r(v)}^{k-1} = \mathbf{AGGREGATE}(h_u^{k-1}, \forall u \in \mathcal{N}_r(v))\\
h^{k}_{v} = \sigma(f( h_v^{k-1}) + g( h_{\mathcal{N}_r(v)}^{k-1} )) \label{updater}\\
f(h_v^{k-1}) = \textbf{W}^k\cdot h_v^{k-1} \label{eq:f}\\
g( h_{\mathcal{N}_r(v)}^{k-1}) =  \textbf{U}^k \cdot\sigma(\textbf{V}^k \cdot h_{\mathcal{N}_r(v)}^{k-1} )\label{eq:g}
\end{eqnarray}
where $h_v^{k}$ denotes the hidden state of node $v$ after the $k$-th convolution layer, and $\mathbf{AGGREGATE}$ is a pooling operator (\textit{mean}, \textit{sum}, and etc.) and $h_{\mathcal{N}_r(v)}^{k-1}$ is the neighborhood vector of $v$ that incorporates the surrounding information of $v$ given a meta relation $r$. Without loss of generality, we instantiate the pooling function as a \textit{sum} operator.
 In \eqref{updater}, $f(\cdot)$ and $g(\cdot)$ are two distinctive data-driven functions that maps $h_v^{k-1}$ and $h_{\mathcal{N}_r(v)}^{k-1}$ to a new feature space respectively. Here $\mathcal{N}_r (v)$ contains the neighbors of $v$ with the top-m largest edge weights. The motivation we select two distinctive functions is due to the heterogeneity of these two hidden representations which correspond to different node types.  For the hidden features of the node itself in \eqref{eq:f}, we directly use a linear projection with a weight matrix $\mathbf{W}^k\in\mathbb{R}^{d\times d}$. For the neighborhood vector in \eqref{eq:g}, we first compress it to a low-dimension vector sized $l$ by a linear transformation with $\textbf{V}^k\in\mathbb{R}^{d\times l}$, where $l<d$,  followed by a activation function $\sigma$. For term convenience, we name $l$ as latent feature size. We then apply another linear combination with $\textbf{U}^k\in\mathbb{R}^{l\times d}$ to map the intermediate hidden features to the original dimension sized $d$. We select such autoencoder-based method for the neighborhood vector because it can be viewed as a noise filter from the neighborhood context which usually contains some irrelevant information. Besides, it can also reduce the computational complexity compared with the direct assignment of a weight matrix sized $d\times d$ on the hidden features from $\mathcal{O}(d^2)$ to $\mathcal{O}(d\cdot l)$.  Although the graph attention network (GAT) \cite{DBLP:conf/iclr/VelickovicCCRLB18,9200731} also addresses the problem to  extract the most influential information from neighbors by the attention mechanism, it has been experimentally verified that GAT performed the same as or worse than GCN in noisy graphs \cite{DBLP:conf/kdd/ZhuZ0019}. This is because GAT introduces too many parameters when learning attention coefficients, which leads to overfitting to noise. Besides, compared with our design, GAT is more computation expensive.

 After feeding the node information to $K$ such layers, which are defined by metapath $p$, we can eventually obtain the node semantic embedding $h_v^p$. To reduce the computational cost, the network parameters of the graph convolution are shared across different nodes at the same layer. 

\vpara{Semantic Attention Layer.} We next introduce the semantic attention layer, which aims to fuse multiple embeddings extracted based on various metapaths. The general idea is as follows: different metapaths reveal different aspects of node context; thus, there is a necessity to aggregate the semantics revealed from different metapaths. To address this issue, we follow the idea proposed in \cite{wang2019heterogeneous}, which uses a self-attention mechanism to capture the diverse semantics revealed from different metapaths. The mathematical expressions of the semantic fusion are as follows:
\begin{eqnarray}
	\tilde{h}_v = \sum_{p\in P} w_{p} \cdot h_v^{p} \quad and \quad
	w_{p} = \frac{exp(W_{att}\cdot h_v^{p})}{\sum_{p\in P}{exp(W_{att}\cdot h_v^{p}})} \label{eq:attn}
\end{eqnarray}
where $w_{p}$ is the learned importance weight of  metapath $p$, which is computed by a scaled self-attention mechanism from \eqref{eq:attn}, and $W_{att} \in \mathbb{R}^{d\times 1}$ is the weight matrix for mapping the original hidden representation $h_v^{p}$ to a scalar which will be scaled to the importance weight of the corresponding metapath $p$.  
\subsection{Siamese Neighbor Matching}
In the past two sections, we have presented the methodology to fuse node-level and subgraph-level information. In an ordinary matching model, the next phase is usually to feed the source and target embeddings into a contrastive loss function, like in \cite{huang2013learning}.  Before calculating the loss function, in this section,  we design a \siamese neighbor matching layer to transform the original matching problem between heterogeneous nodes to a  matching problem between two pairs of  \bidpair, which is illustrated in the left part of \figref{fig:model}. Each pair contains the source or target node and its most influential neighbors. More specifically, the \bidpair pair in the tower of ads contains the embedding of source ad $v_a$ and the averaged embedding of $v_a$'s $\kappa$ most influential neighbors with type of \textit{\bidword}, while the \bidpair pair in the tower of \bidwords contains the embedding of target \bidword $v_q$ and the averaged embedding of its $\kappa$ most influential neighbors with type of \textit{ad}. 
We finally combine the averaged semantic embedding of a node's most influential neighbors with the node itself. 

The motivation here is that although in a heterogeneous matching problem, the embeddings of both the source node (ad) and the target node (keyword) are expected to be close in feature space by optimization, there is no explicit guarantee for such expectations due to their differences in the network structures, parameters, and metapaths to compute the final embeddings. Besides, by introducing so much side-information brought by the heterogeneous graph neural networks, it would become more challenging to satisfy such assumption.  In contrast, compared with only utilizing the source semantic embedding itself, introducing the averaged semantic embedding of its influential neighbors (named neighborhood embedding) would help match with target embedding. This is because they share the same raw feature types, network structures, metapaths, and parameters. Furthermore, the influential neighbors can also directly characterize what the source nodes are like in an e-commerce scenario. The mathematical expressions are formulated as below:
\begin{eqnarray}
z_{v_a}= \tilde{h}_{v_a}  + \mathbb{E}_{u_q \in \mathcal{N}_\kappa(v_a)}\tilde{h}_{u_q} \quad and \quad z_{v_q}= \tilde{h}_{v_q}  + \mathbb{E}_{u_a \in \mathcal{N}_\kappa(v_q)}\tilde{h}_{u_a}  
\end{eqnarray}
where $\mathcal{N}_\kappa(v_a)$ denotes $\kappa$ most influential \bidwords bid by node $v_a$; similarly,  $\mathcal{N}_\kappa(v_q)$ denotes $\kappa$ most influential ads bidding on \bidword $v_q$. In this way, the whole structures to compute $z_{v_a}$ and $z_{v_q}$ are symmetric, which is also the reason we name this network \siamese neighbor matching layer.  It's important to note that  \siamese neighborhood matching mechanism differs from the dual matching scheme used in \cite{DBLP:conf/cikm/LiuGDGGBY20, DBLP:conf/kdd/NiuLLXSDC20, DBLP:conf/kdd/0009ZGZNQH19}. These studies design symmetric network structures for heterogeneous objects in two-tower; however, their neighborhood embedding is computed based on different metapaths and network parameters thus tend to be hard for heterogeneous matching.

\subsection{Multi-view Objectives and View Transformation}
\label{subsec:multiveiw}
For improving the effectiveness of matching with cold-start ads, we introduce different types of relations between ads and \bidwords as our objectives. More specifically, we select click relations between ads and \bidwords, bid relations between ads and \bidwords, and click relations between \bidwords and items (each ad corresponds to an item (\figref{fig:schema})) as our objectives. As different type of objectives might have different feature space to match, we use a view transformation function $f_{trans}(\cdot)$ to map the original embeddings $z_v$ to a view-specific embedding $z^r_v$, where $f_{trans}(\cdot)$ is a view-specific multi-layer perceptron and $r$ represent the view type. 

\subsection{Model Learning}
\label{subsec:model_learn}
We train our \methodshort in a supervised manner using a contrastive loss used in \cite{huang2013learning}.
The basic idea is that we aim to maximize the inner product of positive pairs, i.e., the transformed embedding of the source ad and the corresponding related \bidword.
Meanwhile, we also want the inner product of negative examples, the source ad and its unrelated \bidword, to be smaller than that of the positive sample. To achieve this, we compute the posterior probability of a \bidword given an ad from the semantic relevance score between them
through a softmax function:
\begin{equation}
	p(v_q|v_a, \mathcal{Q}') = \frac{exp(\gamma \cdot z_{v_a} \cdot z_{v_q})}{\sum_{{v_q}'\in \mathcal{Q}'} exp(\gamma \cdot z_{v_a} \cdot z_{{v_q}'})}
\end{equation}
where $\gamma$ is a smoothing factor in the softmax function. $\mathcal{Q}'$ denotes a subset of candidate \bidwords to match. For each positive pair, denoted by ($v_a$, $v_{q+}$) where $v_{q+}$ is the relevant \bidword, $\mathcal{Q}'$ includes $v_{q}^+$ and five randomly selected \bidwords in the candidate set $\mathcal{Q}_{v_a}^c$. The candidate set $\mathcal{Q}_{v_a}^c$ contains the \bidwords that are discriminated by a category classifier as the same category as ad $v_a$.

In the training phase, we optimize the whole networks' parameter set by maximizing the likelihood of the posterior probability of all \bidwords across the training set. Equivalently, we acquire to optimize the following loss function:
\begin{eqnarray}
	L(\Theta)=-\sum_{(v_a, v_q^+, \mathcal{Q}')}p(v_q^+|v_a, \mathcal{Q}')
\end{eqnarray}
where $\Theta$ represents the parameter set of the whole networks which is updated iteratively by an Adam optimizer.

\subsection{Deployment}
\vpara{Pipeline of Keyword Recommendation. }Keyword recommendation is a multi-stage system, consisting of matching, filtering and ranking phases. In the keyword matching phase,  we first handle network construction from user behavioral records and advertiser bidding records offline by MapReduce downstream in the Alibaba Cloud platform\footnote{https://us.alibabacloud.com}.  During the training and inference, we leverage the distributed deep learning framework XDL\footnote{https://github.com/alibaba/x-deeplearning} and graph search engine, Euler\footnote{https://github.com/alibaba/euler}, to conduct run-time subgraph extraction and parameter update in a distributed fashion.  After inference, ad embeddings and keyword embeddings are stored in a database for online services, and we use an approximate nearest neighbor (ANN) search engine to retrieve the most relevant keywords for each ad. In the filtering phase, we use a term-match based relevance model to filter the unrelated keywords for each ad. Lastly, in the ranking phase, we use an MLP-based model with enriched features to estimate the potential number of clicks brought by each remained keyword and rank those keywords based on the estimated values. In this way, we can recommend advertisers with the keywords that can bring as many user clicks as possible.

\vpara{Acceleration.}  \methodshort can be implemented by sampling subgraphs separately for each node and computing each node's corresponding final embedding via neural networks, which is computation consuming. As the extracted subgraphs based on the top-k sampling strategy of a particular node and its influential neighbors usually share plenty of relation-paths, we lighten the computation burden by only calculating the intermediate embedding once for each distinctive relation-path.

\vpara{Negative Sampling.}
As illustrated in \secref{subsec:model_learn}, for each positive pairs, we need to sample multiple negative keyword nodes for training. The whole procedure for negative sampling can be divided into two phases: 1) we calculate the weights (square root of the searched count) of keywords in each leaf category offline. 2) At run time of training, for each positive \bidpair, we pick five negative keywords in the same leaf category by weighted random selection with the help of the graph index engine Euler.
\hide{
In this section, we present the deployment of the proposed \bidword matching model in \zhitongche.  The construction of the heterogeneous information graph is handled offline by MapReduce downstream in Alibaba Cloud platform\footnote{https://us.alibabacloud.com}, and both training and inference components are implemented via XDL framework\footnote{https://github.com/alibaba/x-deeplearning}. Rather than extracting subgraphs offline for every source or target node, we fetch nodes' neighborhoods in the run time of training or inference from the graph indexing engine Euler\footnote{https://github.com/alibaba/euler} by hash keys. The inference component is very similar to the training component except without backward propagation. After inference, ad embeddings and \bidword embeddings are stored in a database for online/offline service, and we use an 
approximate nearest neighbor (ANN) search engine to retrieve the most related \bidwords for each ad. 
}
\hide{
	 \begin{figure}[t]
		\centering
		\includegraphics[width=0.46\textwidth]{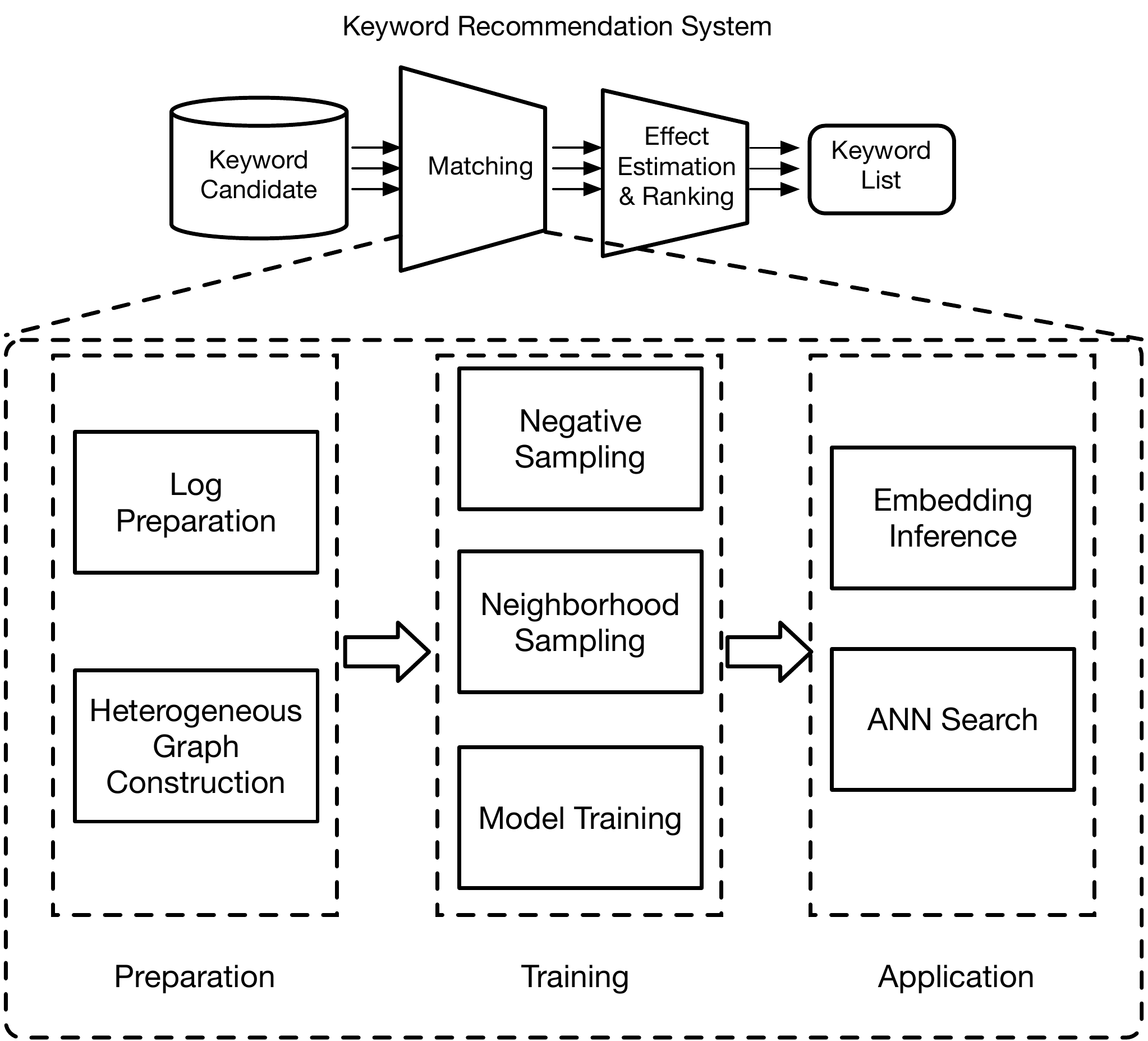}
		\caption{Application of \bidword matching in \zhitongche.}
		\label{fig:app}
	\end{figure} 
}

\section{Experiments}

\subsection{Experimental Setup}
\vpara{Dataset Description.}
We evaluate our proposed method based on a real-world dataset collected from the search logs in Taobao platform and the advertiser behavioral records in \zhitongche. The dataset covers seven consecutive days of these records, spanning from August 19th, 2020 to August 26th, 2020. We construct our heterogeneous network  from these logs based on the network schema illustrated in \figref{fig:schema}, where the edge weight (the importance of an edge) is defined as the appearing times of a particular relation in the records. For the training set, we sample 10M ad-keyword pairs in each view. The target relation set contains 50M click relations between ads and keywords in the constructed network. To prevent information leakage, we drop the relations in the target relation set from the original network. The more specific data statistical information is exhibited in \tableref{tab:dataset}. 

\begin{table}[ht!]
	\centering
	\small
	\caption{Statistics of the Dataset.}
	\renewcommand{\arraystretch}{\tablestrech}
	
	\begin{tabular}{l|ccccc}
		\toprule
		Relation(A-B)& \#A & \#B & \#A-B & \#labeled A-B & \#target A-B  \\
		\midrule
		a\_\textit{click}\_q &  50M  & 10M & 500M & 10M & 50M  \\
		a\_\textit{bid}\_q &  50M  & 10M & 5B & 10M & -\\
		i\_\textit{click}\_q &  100M & 100M  & 10B & 10M & - \\
		a\_\textit{coclick}\_i &  5M  & 50M & 500M & - & - \\
		\bottomrule
	\end{tabular}
	\hide{
	\begin{tabular}{l|ccc}
		\toprule
		Relation(A-B)& \#A-B & \#labeled A-B & \#target A-B  \\
		\midrule
		a\_\textit{click}\_q &  100M & 10M & 100M  \\
		a\_\textit{bid}\_q &  1B & 10M & -\\
		i\_\textit{click}\_q &  1B & 10M & - \\
		a\_\textit{coclick}\_i & 100M & - & - \\
		i\_\textit{coclick}\_i  & 1B & - & -\\
		\bottomrule
	\end{tabular}
	}
	\label{tab:dataset}
\end{table}
\vpara{Baselines.}
To demonstrate our proposed method's effectiveness, we compare our method with multiple baseline methods and their variants. To ensure the fairness of the comparison, the results we reported are all under the multi-view learning framework mentioned in \secref{subsec:multiveiw} unless otherwise stated. 

\begin{itemize}
    \item \textbf{Term-Match}: This is a \bidword retrieval method that extracts related \bidwords by calculating the term similarity between ads' titles and \bidwords, which is one of the \bidword matching models currently deployed in \zhitongche.
	\item \textbf{DSSM}: This is a two-tower matching model that projects the features of  objects to a low dimensional space via multi-layer perceptrons (MLPs) \cite{huang2013learning}.
	\item \textbf{HAN}: We replace the MLPs in DSSM by heterogeneous attention networks (HANs) proposed in \cite{DBLP:conf/www/WangJSWYCY19}. The metapaths to aggregate neighbors are mentioned in \secref{sec:subgraph}.
	\item \textbf{IntentGC}: This is a dual-HGCN-based recommendation model that captures heterogeneous relations between nodes with the same node types \cite{DBLP:conf/kdd/0009ZGZNQH19}.
	\item \textbf{\methodshort}: This is our proposed model. We set the embedding length $d$ as 64 and the latent size $l$ as 16. The model is optimized using an Adam optimizer with a learning rate of 0.03, and the batch size is 512. 
	\item \textbf{DSSM(s), HAN(s), IntentGC(s)}: Based on DSSM, HAN or IntentGC, we add the \siamese matching layer between the output of multi-layer perceptrons and the multi-view transformation layer's input.
	\item \textbf{\methodshort$_{\backslash s}$}: This model drops the \siamese neighbor matching layer in \methodshort. The remaining settings for this variant are the same as for our proposed methods.
	\item \textbf{\methodshort$_{\backslash v}$}: This model only utilizes click data between ads and \bidwords as the labeled positive samples.
	\item  \textbf{\methodshort$_{\backslash a}$}: This variant replaces the graph convolution layer with autoencoder of \methodshort by GraphSage \cite{hamilton2017inductive}.
	\item \textbf{\methodshort$_{bid}$,~\methodshort$_{item}$}: Each of these variants only use the bid/item relations to aggregate neighbors' information.
\end{itemize}

\begin{table}[ht!]
	\vspace{-0.02in}
	\label{tab:results_problems}
	\centering
	\small
	\vspace{-0.03in}
	\caption{Performance comparison. }
	\renewcommand{\arraystretch}{\tablestrech}
	\begin{tabular}{l|llll}
		
		\toprule
		K& 100 & 200 & 500 & 1000\\
		Method &  \multicolumn{4}{c}{Recall@3K}  \\
		\midrule
		Term-Match & 10.74\%  & 12.79\%  & 16.10\%  & 16.74\% \\
		\midrule
		DSSM & 13.73\%  & 23.15\%  & 41.34\%  & 57.74\% \\
		HAN &  16.72\%  & 27.65\%  & 47.91\% &  64.83\% \\
		IntentGC & 16.97\% & 26.63\% & 43.57\%  & 57.04\% \\
		\midrule
		DSSM(s) & 16.45\%  & 27.24\% & 47.32\%  & 64.35\% \\
		HAN(s) & 17.52\% & 28.82\% & 49.75\%  & \textbf{67.33\%} \\
		IntentGC(s) & 17.73\% & 27.93\% & 45.68\%  & 60.01\% \\
		\midrule
		\methodshort$_{\backslash a}$  & 18.82\%  & 29.60\% & 48.34\%  & 63.31\% \\
		\methodshort$_{\backslash s}$  & 19.32\%  & 30.28\%  & 49.77\% & 65.63\% \\
		\methodshort$_{\backslash v}$ &  17.99\%  & 23.91\% & 26.66\%  &47.43\% \\
		\methodshort & \textbf{19.82\%}  & \textbf{30.93\%}  & \textbf{50.49\%}  & 67.16\% \\
		\midrule
		\methodshort$_{bid}$ & 15.97\% & 26.74\%  & 46.33\%  & 64.81\% \\
		\methodshort$_{item}$ & 15.78\% & 26.62\% & 47.48\% & 66.11\% \\
		\bottomrule
	\end{tabular}
\vspace{-0.05in}
	\label{tb:performances}
\end{table}

\begin{table*}[ht!]
	\label{tab:results_problems}
	\centering
	\small
	\caption{Performance comparison in different views.}
	\renewcommand{\arraystretch}{\tablestrech}
	\begin{tabular}{l|cccc|cccc|cccc}
		\toprule
		View & \multicolumn{4}{c|}{Recall@K (ad-click)} & \multicolumn{4}{c|}{Recall@K (item-click)} & \multicolumn{4}{c}{Recall@K (ad-bid)}\\
		K& 100&200&500&1000 & 100&200&500 &1000&100&200& 500 & 1000\\
		\midrule
		DSSM &  4.52\%  & 7.71\%  & 14.46\% &  22.40\% &  4.63\%  & 7.87\%  & 14.61\% &  22.26\% & 4.46\% & 7.68\% & 14.18\% & 22.03\% \\ 
		HAN & 5.83\% &  9.90\%  & 18.38\%  & 27.47\% & 5.78\% &  9.80\%  & 18.30\%  & 27.44\% & 5.80\% &  9.81\%  & 18.06\%  & 26.69\%\\
		IntentGC & 6.33\% &  10.23\%  & 18.40\%  & 27.42\% & 6.16\% &  10.02\%  & 18.19\%  & 27.34\% & 6.13\% &  10.09\%  & 18.34\%  & 27.31\%\\
		\methodshort$_{\backslash v}$ &  4.51\% &  7.67\%  & 14.72\%  &24.39\% &  - &  -  & -  & - &  - &  - & -  &-\\
		\methodshort & \textbf{7.07\%}  & \textbf{11.21\%}  & \textbf{19.75\%}  & \textbf{28.99\%} & \textbf{7.00\%}  & \textbf{11.14\%}  & \textbf{19.70\%}  & \textbf{28.96\%} & \textbf{6.45\%}  & \textbf{10.52\%}  & \textbf{18.84\%}  & \textbf{27.82\%}\\
		\midrule
		\methodshort$_{bid}$ & 5.58\% & 9.50\% & 18.00\% & 27.55\%  & 5.47\% & 9.34\% & 17.74\% & 27.33\% & 5.49\% & 9.41\% & 17.70\% & 26.68\%\\
		\methodshort$_{item}$ & 5.53\% & 9.41\% & 17.71\% & 26.77\%& 5.61\% & 9.57\% & 17.98\% & 27.10\% & 5.44\% & 9.31\% & 17.52\% & 26.30\%  \\
		\bottomrule
	\end{tabular}
	\label{tb:multi_channel_evaluation}
\end{table*}


\subsection{Performance Evaluation}
\label{subsec:performance_evaluation}
As our task focuses on retrieving hundreds of \bidword candidates, we use the recall rate as our evaluation metrics as illustrated in \secref{sec:setup}, which is also adopted in \cite{huang2020embedding}. As our model is under a multi-view framework (the number of view is 3), we use Recall@3K instead of Recall@K as our metrics. For the approaches under the multi-view framework, we retrieve each ad's the top-K most related \bidwords in each view, and aggregate these relations as the final retrieved set $\mathcal{O}^{3K}$ to compute Recall@3K. To ensure fair comparison, for those methods that are not under a multi-view framework (Term-Match and \methodshort$_{\backslash v}$), we directly retrieve the top-3K most related \bidwords for each ad to compute Recall@3K.

\tableref{tb:performances} reports the performance of our proposed method compared to other competing methods or some variants from them. It can be seen from the table that our model can consistently outperform all comparative approaches or variants by achieving the highest recall rate in most cases. Firstly, we compare the relevance-based model deployed in our platform with other embedding-based retrieval methods. We observe that our relevance-based model (Term-Match) cannot achieve a good performance, especially when K is large. This finding demonstrates that the retrieval model that only relies on text relevance might not have good practical performance. Then we compare \methodshort with  DSSM that does not utilize neighbor information. We note that our method achieves much better performance, as \methodshort can capture auxiliary information from heterogeneous relationships. Besides, we compare \methodshort with GNN-based methods (HAN, IntentGC); here, HAN and IntentGC also performs the aggregation operations on neighbors, but it does not explicitly consider the misleading influence brought by noisy links and features, and the attention mechanism used in HAN's aggregator might lead to overfitting.

To further demonstrate how our \siamese neighbor matching layer helps matching task, we conduct several ablation studies on DSSM(s), HAN(s), IntentGC(s) and \methodshort$_\backslash s$. The comparison between DSSM with DSSM(s), HAN with HAN(s), IntentGC with IntentGC(s) and \methodshort with \methodshort$_{\backslash s} $ presents the effectiveness of introducing \siamese neighbor matching in a DNN-based matching task. Interestingly, we observed that adding our \siamese neighbor matching layer to a simple DNN-based matching approach (DSSM) can significantly improve performance, achieving very close performance to other GNN-based methods. This again shows the importance of conducting \siamese neighbor matching. We can also note that  \methodshort surpasses the performance of \methodshort$_{\backslash a}$. One reason for this is that the autoencoder based graph convolution layer prevents misleading information from propagating from neighbors and thus enhances the quality of neighbor embeddings. Besides, by eliminating the multi-view framework, we observe a drastic drop in performance, which demonstrates the need to involve different types of objectives in learning.

We also explore how our proposed model performs with different groups of metapaths. Compared with sampling neighbors based on both bid-based and item-based metapaths, sampling neighbors with only one group of metapath harms the performance. This is because the bid-based and item-based relations can provide useful information from different aspects, which can complement each other. It would also be interesting to note that the performance of \methodshort$_{bid}$ is better than the performance of \methodshort$_{item}$ when K is no more than 200, and vice versa when K is more than 200. This is because the item-based metapaths brings more diverse neighbors compared with those sampled based on the bid-based metapaths (since this group is not directly related to the object task), which might hamper the quality of matched elements whose matching score is relatively high, but improve the total recall rate when K is relatively large. This also shows the importance of selecting proper types of metapaths in the task of \bidword matching.

\subsection{Performance Evaluation in Different Views}
This section aims to compare the performance of different approaches for each view objectives (\tableref{tb:multi_channel_evaluation}). Our model achieves state-of-the-art performance in each view, and its improvement on the performance over other baseline methods is significant. This finding demonstrates our model can achieve a good performance not only in total but also in each channel. Another interesting finding is that we note that the performance in the channel of bidding relations is worse than that in the channels of item-clicking and ad-clicking in most cases. This is because the labels of bidding relations are extracted by advertisers' manual selection of \bidwords, which might introduce some non-professional bidding behaviors and achieve performance. Besides, we also note that \methodshort achieves better performance over \methodshort$_{\backslash v}$ in the view of ad-click, which demonstrates that learning with auxiliary labels in other views can improve the performance in the each view.

\subsection{Performance Evaluation on Cold Start Ads}
Conducting cold-start recommendation is a critical problem for a recommender system, which is not an exception for \bidword recommendation. In this section, we present the experimental results for the cold-start ads. More specifically, \tableref{tb:cold start performance} reports the performance of different methods on the ads which are newly created on August 26, 2020. Unlike the target set we used in  \secref{subsec:performance_evaluation}, we select the candidate set of a new ad as the \bidwords that bring user clicks to the ad in the next 14 days. The performance of these methods is consistent with that in \secref{subsec:performance_evaluation}. \methodshort performs the best across all the competitive approaches, which demonstrates the \methodshort works well on the whole dataset and has an excellent capability to retrieve useful \bidwords for new ads. More specifically, we find that learning with various groups of metapaths and using  autoencoder module in graph convolution and \siamese neighbor matching layer can improve the performance on cold-start ads.

\begin{table}[H]
	\label{tab:results_problems}
	\centering
	\small
	\vspace{-0.05in}
	\caption{Performance comparison of cold-start ads.}
	\renewcommand{\arraystretch}{\tablestrech}
	\begin{tabular}{l|llll}
		\toprule
		K& 100 & 200 & 500 & 1000\\
		Method &  \multicolumn{4}{c}{Recall@3K}   \\
		\midrule
		Term-Match & 15.14\% &17.10\% &18.82\% & 19.86\%\\
		\midrule
		DSSM &  13.02\%  & 21.90\%  & 38.84\% &  53.82\% \\
		HAN & 17.29\% &  28.75\%  & 50.21\%  & 68.02\% \\
		IntentGC & 16.01\% & 25.22\% & 41.41\%& 54.26\%\\
		\midrule
		\methodshort $_{\backslash a}$ &  17.05\% &  28.13\%  & 48.93\%  &67.01\% \\
		\methodshort $_{\backslash s}$  & 19.32\%  & 30.28\%  & 49.77\%  & 65.63\% \\
		\methodshort $_{\backslash v}$ &  12.95\% &  22.91\%  & 31.78\%  &41.50\% \\
		\methodshort & \textbf{20.50\%}  & \textbf{32.25\%}  & \textbf{53.22\%}  & \textbf{70.40\%} \\
		\midrule
		\methodshort$_{bid}$ & 16.35\% & 27.58\% & 49.35\% & 68.77\% \\
		\methodshort$_{item}$ & 16.22\% & 27.33\% & 48.99\% & 68.31\% \\
		\bottomrule
	\end{tabular}
	\label{tb:cold start performance}
	\vspace{-0.05in}
\end{table}

\subsection{Online Evaluation}
\label{subsec:abtest}
To evaluate the effectiveness of our proposed method on the real-world products, we deployed our \methodshort both on the \bidword suggestion tool and the automatic \bidword hosting tool of \zhitongche. More details about these tools can be referred to in the part of Appendix. 
\tableref{tab:online} presents the results of online AB tests on the \bidword suggestion tool and the automatic \bidword hosting tool in \zhitongche. For the \bidword suggestion tool, our proposed method achieves an improvement of +4.19\% in terms of the adopting rate and +5.35\% in terms of the number of clicks over the online deployed term-match based model. For the automatic \bidword hosting tool, our proposed method obtains an improvement of +10.89\% in terms of the number of clicks over the online deployed GraphSage-based matching model.
\begin{table}[H]
	\centering
	\small
	\caption{Online performance of compared methods in different scenes.}
	\renewcommand{\arraystretch}{\tablestrech}
	\begin{tabular}{l|l|l}
		\toprule
		Scene & Baseline & Lift   \\
		\midrule
		\multicolumn{1}{p{2.5cm}|}{\Bidword suggestion tool }&\multicolumn{1}{p{2.4cm}|}{a term-match-based model } &   \multicolumn{1}{p{2.5cm}}{adopting rate + 4.19\%, \#click +5.35\%}  \\
		\midrule
		\multicolumn{1}{p{2.5cm}|}{\Bidword hosting tool} &\multicolumn{1}{p{2.4cm}|}{a  two-layer GraphSage model}& \#click +10.89\%  \\
		\bottomrule
	\end{tabular}
	\label{tab:online}
\end{table}

\section{Conclusion}
\label{sec:conclusion}
In this paper, we proposed \methodshort for the \bidword matching problem based on a metapath-based heterogeneous GNN, which consists of a hierarchical structure to capture complex structures and rich semantics behind heterogeneous information networks in a robust way. The proposed model leverages node-level fusion, subgraph-level fusion, and \siamese neighbor matching to adaptively aggregate relevant neighborhood patterns.  By introducing the autoencoder-based graph convolution layer and the \siamese matching layer, \methodshort can mitigate the negative effect from noisy patterns and enhance relevant information.  In addition, we use a multi-view framework to learn the posterior probability of various objectives, which helps introduce more supervised signals. Experimental results show that our model consistently outperforms competitive approaches. Extra online AB tests also demonstrate the superiority over existing  methods deployed online. 

For future work, we'd like to continue exploring into other sophisticated relations hidden in the heterogeneous graph, such as text similarity, user profiles, etc. Furthermore, \methodshort still relies on human-crafted metapaths while recently there are a few transformer-based models that could potentially learn all aggregation strategies by themselves \cite{10.1145/3366423.3380027}, which is an auspicious direction. Last, considering Bert-like pre-trained models have pushed state of the art in many areas of NLP, combining Bert with GNN to enhance the quality of keyword retrieval may also be promising. That being said, both transformer-based HGNN and Bert-like models are usually of very large model size, which would be a great challenge for training and deployment in industrial systems. Further efforts like model compression might also be required.
\section*{appendix}

\subsection*{Hyperparameter Setting} For the implementations of all comparable methods, we set the same hyperparameters. 
\begin{table}[ht]
	\centering
	\vspace{-0.04in}
	\renewcommand\arraystretch{\tablestrech}
	\small
	\caption{Hyperparameter settings.}
	\begin{tabular}{l|c|c|c}
		\toprule
		\textbf{Hyperparameter} &value &\textbf{Hyperparameter} &value \\ 
		\midrule
		\textbf{learning rate} & 0.03 & \textbf{optimizer} & Adam \\ 
		\textbf{mini-batch size} & 512& \textbf{\#epoch}  & 5 \\ 
		\textbf{latent feature size $l$} &16& \textbf{hidden size $d$}  &64\\
		\textbf{\#neighbor $m$} &10 & \textbf{\#\siamese neighbor $\kappa$}&3    \\ 
		\textbf{\#negative samples}&5&  &  \\ 
		\bottomrule
	\end{tabular}
	\vspace{-0.1in}
\end{table}

\subsection*{Details of Advertiser Tools for Online AB tests}

\vpara{\Bidword Suggestion Tool.} This is a suggestion tool for advertisers when they manually add \bidwords. More than 30\% of new \bidwords with impressions are brought by the suggestion tools each day. For each ad, the suggestion tool will present hundreds of \bidword candidates; the advertisers can adopt the candidates from the suggestion list.  

The control group contains a relevance-based \bidword matching model. In the treatment group, we combine \bidwords retrieved by \methodshort and the relevance-based model. The adopting rate of the suggested \bidwords and the total clicks brought by the suggested \bidwords are the main metrics that we are concerned about.

\vpara{\Bidword Hosting Tool.} In \zhitongche, advertisers are provided with many black-box marketing tools such as the automatic \bidword hosting tool. By merely setting budgets and bid prices, advertisers can handle online marketing for sponsored search without explicitly choosing any \bidwords. The automatic hosting procedure hidden behind the product can be divided into two parts. The first part is the addition of \bidwords based on the retrieved \bidwords with estimated ad clicks, while another part is to delete existing \bidwords that cannot bring impressions or have very low ctr/cvr. 

In the experiment, the control group contains a two-layer GraphSage based matching model, which is proven effective in the previous AB test. In the treatment group, we use the \bidwords retrieved by \methodshort. We use the averaged ad clicks brought by daily \bidword addition to evaluate different methods' performance.

\bibliographystyle{ACM-Reference-Format}
\balance
\bibliography{reference}

\end{document}